\documentclass[12pt]{iopart}

\expandafter\let\csname equation*\endcsname=\relax 
\expandafter\let\csname endequation*\endcsname=\relax 
\usepackage{amsmath,amssymb,amsthm} 

\usepackage{graphicx} 
\usepackage{xcolor}

\begin{document}

\title[A numerical damped oscillator approach to constrained Schr\"{o}dinger equations]{A numerical damped oscillator approach to constrained Schr\"{o}dinger equations}

\author{M \"{O}gren}

\address{School of Science and Technology, \"{O}rebro University, 701 82 \"{O}rebro, Sweden,}
\address{Hellenic Mediterranean University, P.O. Box 1939, GR-71004, Heraklion, Greece.}
\ead{magnus.ogren@oru.se}

\author{M Gulliksson}

\address{School of Science and Technology, \"{O}rebro University, 701 82 \"{O}rebro, Sweden,}
\address{Institutt for data og realfag, H\o gskulen p{\aa} Vestlandet, 5020 Bergen, Norway.}
\ead{marten.gulliksson@oru.se} 
\


\vspace{10pt}

\date{\today}

\begin{abstract}
This article explains and illustrates the use of a set of coupled dynamical equations, second order in a fictitious time, which converges to solutions of stationary Schr\"{o}dinger equations with additional constraints.
In fact, the method is general and can solve constrained minimization problems in many fields.
We present the method for introductory applications in quantum mechanics including three qualitative different numerical examples: the radial Schr\"{o}dinger equation for the hydrogen atom;
the two-dimensional harmonic oscillator with degenerate excited states;
and a nonlinear Schr\"{o}dinger equation for rotating states.
The presented method is intuitive, with analogies in classical mechanics for damped oscillators, and easy to implement, either in own coding, or with software for dynamical systems.
Hence, we find it suitable to introduce it in a continuation course in quantum mechanics or generally in applied mathematics courses which contain computational parts.
The undergraduate student can for example use our derived results and the code (supplemental material) to study the Schr\"{o}dinger equation in 1D for any potential.
The graduate student and the general physicist can work from our three examples to derive their own results for other models including other global constraints.
\end{abstract}
%
\vspace{2pc}
\noindent{\it Keywords}: Schr\"{o}dinger equation, eigenenergies, degeneracy, nonlinear Schr\"{o}dinger equation, constraints
%
%
%
%

\section{Introduction}

In this article we describe the idea of solving stationary Schr\"{o}dinger equations (SE) as energy minimization problems with constraints, by using a second order damped dynamical system.
We discuss how to numerically solve the problems in a stable and efficient way. 

For the formulas in this article to be easily recognized and directly applicable for the students in different courses, we write most formulas explicitly in an infinite-dimensional setting. 
For example, we use integrals instead of scalar products (or Dirac notation). However, if you write your own code~\cite{program} instead of using high-level solvers for the differential equations, you need to formulate integrals as finite sums and derivatives, e.g., as finite differences, i.e., the linear Schr\"{o}dinger equation can be formulated as a linear eigenvector equation $Hu = Eu$, with $H$ a matrix, $u$ a (column) eigenvector, and $E$ an eigenvalue of $H$.

We provide enough details, including the references, for all the numerical results presented here to be reproducible.
We recommend anyone who wants to use the method in practice, or who just wants to obtain a step-by-step understanding of the algorithm, to read (and run) the provided carefully implemented and commented codes~\cite{program}.

From a pedagogical point of view, the method has the advantage of being able to solve a large set of problems in arbitrary dimensions with the same main idea.

The novel way we formulate constraints here allows for using high-level software instead of your own code, for example when calculating excited states. 

There are of course many other ways of attaining the stationary solution of the SE numerically. 
For the common, so called imaginary time dependent SE (se below), the approach is after discretization in space (finite differences, finite elements, or other methods) to solve a first order damped time dependent equation numerically, see~\cite{Schroeder_ajp_2017, SMYRLIS2004436}. 
Sometimes these methods are called steepest descent methods (not to be confused with steepest descent methods in optimization~\cite{NoceWrig06_Numerical_Optimization}). 
In this category we also have the so called shooting methods~\cite{ref_shooting_methods}, although restricted to systems in one spatial dimension or with potentials obeying separation of variables. 
These are in general computationally expensive, but have some advantages for problems with complicated boundary conditions, and for problems with discontinuous solutions.

When the SE is independent of time (e.g. through separation of variables) the SE is, after discretization in space, equivalent to a (nonlinear) finite dimensional minimization problem, with nonlinear constraints.  
Generally, such minimization problems can be solved by a variety of numerical methods including gradient descent methods, (Quasi-) Newton methods, machine learning techniques etc.~\cite{NoceWrig06_Numerical_Optimization, sra2012optimization}. 
Note that in the linear case with only normalization constraints we have a linear eigenvalue problem solved, e.g., by so called diagonalization~\cite{ref_diagonalization_methods}, for which there are numerous specialized numerical methods.  

Our method presented here can be used in a very general and efficient way to minimization problems (linear, nonlinear, with any type of global constraints) and has been shown to be highly competitive~\cite{Gulliksson_book_chapter_2019, Baravdish2019SvenssonGullikssonZhang, SandinOgrenGulliksson2016}, also for linear eigenvalue problems \cite{Gulliksson2017, GullikssonEdvardssonLind2013}.

We hope the readers will expand the theory and applications in different directions from the examples presented here.

\section{The Method}
\begin{figure}[!htb]
\hspace{24mm} 
\includegraphics[scale=0.4]{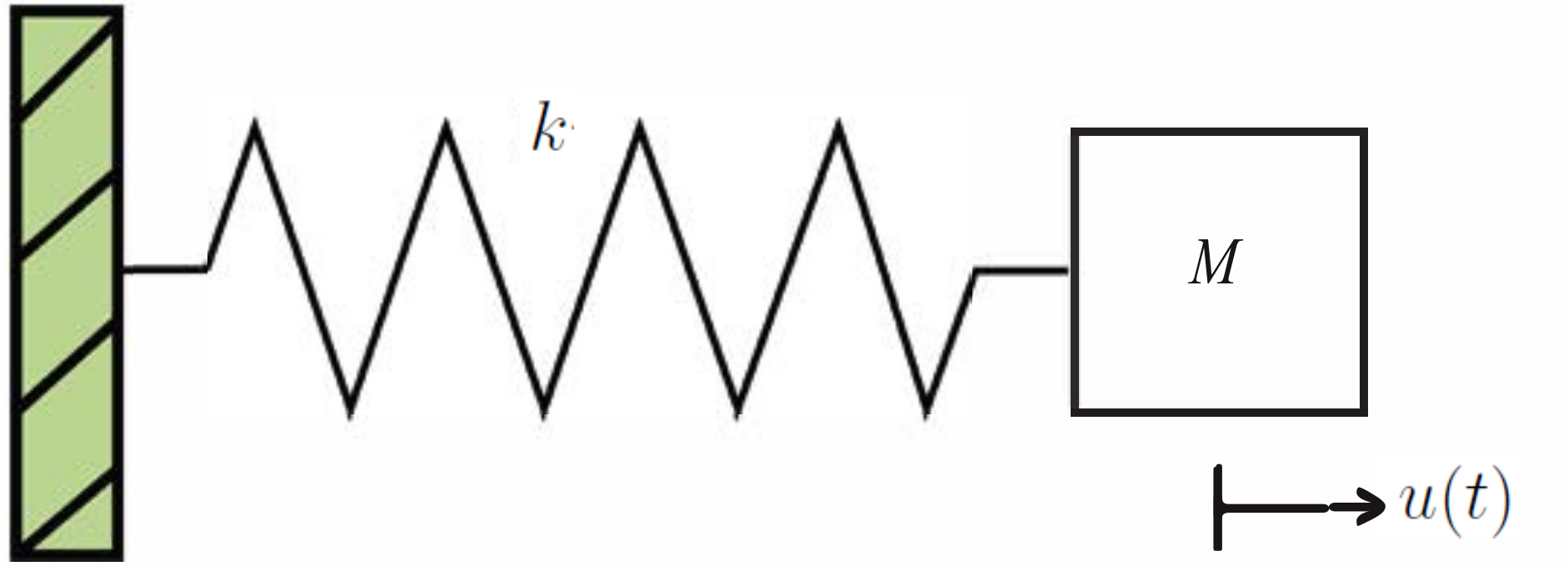}
\caption{A simple oscillating spring-mass system.}
\label{fig:SpringMass}
\end{figure}
\noindent
In order to introduce the idea of the method let us first consider a basic example in classical mechanics.
The  harmonic oscillator is according to Newton's second law described by
\begin{equation}\label{eq:harmosc}
M\ddot{u} + ku = 0, \quad k>0.
\end{equation}
Here  $u=u(t)$ is the distance from the equilibrium for a mass $M$ on which a force $F=-ku$ is acting.
The dot denotes the time derivative. 
For example the
mass could be attached to a spring, with $k$ being the spring constant, see Fig.~\ref{fig:SpringMass}.
If we in addition assume that there is some linear resistance proportional to the velocity $\dot{u}$, e.g.,  between the mass and the surface on which it is
sliding, we get a damped second order system
\begin{equation}\label{eq:dharmosc}
M\ddot{u} + \eta \dot{u} - F(u) = 0,
\end{equation}
where $\eta>0$ is a damping parameter due to the linear resistance.  
With the damping term present the kinetic energy will decay with time, such that $ M\ddot{ u } + \eta \dot{ u } \rightarrow 0 $ when $ t \rightarrow \infty $.
It is clear from this example that one of the parameters $M, \: \eta, \: k$ can be scaled out, so we usually set $M=1$ in the following.
The solutions of Eq.~(\ref{eq:dharmosc}) in the non-critical case ($\eta \neq 2 \sqrt{k}$) are given by
\begin{equation}
u(t) = C_1 \exp(\xi_1 t) + C_2 \exp(\xi_2 t), \label{eq:osc_solutions}
\end{equation}
where $C_i$ are determined by the initial conditions $u(0)$ and $\dot{u}(0)$, and
$\xi_j = -\eta/2 \pm \sqrt{(\eta/2)^2 - k}.$ 
It is easy to see from Eq.~(\ref{eq:osc_solutions}) that $u$ tends to zero when $t$ goes to infinity, which is the equilibrium position of the mass and hence the stationary solution to Eq.~(\ref{eq:dharmosc}).
The value $\eta=2 \sqrt{k}$ for the damping parameter ensures the $\xi_j$ to be real and is referred to as critical damping for Eq.~(\ref{eq:dharmosc}), for which $u(t) = (C_1 + C_2 t) \exp(- \sqrt{k} t)$.
For smaller or larger values of $\eta$, the oscillations are referred to as under- or over-damped respectively.
The critically damped system is known to be the fastest way for the system to return to its equilibrium, i.e., to reach the stationary solution.
In multimode discretized systems the above argument can be generalized in order to obtain an optimal value for the damping parameter to be used in a numerical calculation~\cite{Gulliksson_book_chapter_2019}.

Now we note that $u=0$ is also the solution of the (trivial) minimization problem $\min \left( ku^2/2 \right)$.
In  fact, the convex functional $V(u) = ku^2/2$ is the mechanical potential corresponding to the force $F = -ku$ which is conservative, i.e., $F = -dV/du$. 
Of course, in this case it is easier to directly solve $ku=0$ or $\min ku^2/2$ for $u$, than to integrate the differential equation~(\ref{eq:dharmosc}). 
However, this simple idea can be extended to solve more challenging problems, 
where $\eta$ takes the role of a parameter that can be tuned for optimal numerical properties,
as we explain in this article.

\subsection{Damped oscillator approach to the groundstate of the Schr\"{o}dinger equation} 

We here briefly repeat how the stationary (i.e., time-independent) SE for a particle with mass $M$ and a spatially-dependent potential $V(\mathbf{r},t) = V(\mathbf{r})$ is attained from the time-dependent~SE 
\begin{equation}
\label{eq:TDSE}
   \mathrm{i} \hbar \frac{ \partial \Psi}{\partial t} = - \frac{\hbar^2}{2M} \nabla^2 \Psi + V \Psi \equiv \hat{H} \Psi.
\end{equation}
Given a time and space separating ansatz of the wavefunction $\Psi(\mathbf{r},t)= u(\mathbf{r}) \exp( - \mathrm{i} E t/\hbar )$, we then have from Eq.~(\ref{eq:TDSE})
\begin{equation}
\label{eq:SE}
  \hat{H} u =  Eu.
\end{equation}

Alternatively, the stationary SE can be viewed as the Euler-Lagrange equation~\cite{Calculus_of_variation_book} corresponding to the minimization of the energy, which, if we start from Eq.~(\ref{eq:SE}), is the following functional of~$u$
\begin{equation}
\label{eq:E_u_ratio}
E(u) = \frac{ \int \bar{u} \hat{H} u \, d \mathbf{r} }{ \int |u|^2 d \mathbf{r} },
\end{equation}
where the bar from now on denotes complex conjugation.

If the normalization of the wave function is considered as a constraint, then the denominator of Eq.~(\ref{eq:E_u_ratio}) is unity and the groundstate of Eq.~(\ref{eq:SE}) is given by the solution of
\begin{equation}
\label{eq:E_u}
E= \min_u  \int \bar{u} \hat{H} u \, d \mathbf{r}, \ \textrm{s.t.} \ \int |u|^2 d \mathbf{r} = 1, 
\end{equation}
where \textit{s.t.} is an abbreviation for \textit{subject to}. 
We later give examples with more complicated constraints.


Our main idea for solving Eq.~(\ref{eq:E_u}) is to mimic Eq.~(\ref{eq:dharmosc})
and consider the second order damped dynamical system 
\begin{equation}
\label{eq:dynamgV}
    \ddot{u} + \eta \dot{u} +  \frac{\delta E}{\delta \bar{u}} = 0,\, \eta>0,  \ \int |u|^2 d \mathbf{r} = 1.
\end{equation}
In the following we reserve the dot notation for the derivatives with respect to a fictitious time $\tau$.
The force $F$ in Eq.~(\ref{eq:dharmosc}) corresponds to the generalized force $-\delta E / \delta \bar{u}$, the third term in Eq.~(\ref{eq:dynamgV}),
i.e., to a functional derivative of the energy.
It can be shown~\cite{Begout2015} that the stationary solution to Eq.~(\ref{eq:dynamgV}), say $u^*(\mathbf{r})$, is the solution to Eq.~(\ref{eq:E_u}). 
Note that in the limit when $\tau \rightarrow \infty$ the stationary solution $u^*$ will satisfy the Euler-Lagrange equation $\delta E / \delta \bar{u} = 0$.
The corresponding energy is $E(u^*) = \min_u \left( \int \bar{u} \hat{H} u \, d \mathbf{r} \right)$ where $\hat{H}$ is the Hamiltonian operator from the SE~(\ref{eq:TDSE}).
After the problem has been formulated as in Eq.~(\ref{eq:dynamgV}), an important question is the following: 
How do we choose a stable and efficient numerical method for obtaining the stationary solution to Eq.~(\ref{eq:dynamgV})? 
Symplectic integration methods~\cite{HairerLubichWanner2006} are tailor-made for Hamiltonian systems.
This serves as the motivation for our choice of numerical method.
Let us rewrite Eq.~(\ref{eq:dynamgV}) as the first order system
\begin{equation}
\label{eq:uvSystemVu}
		\begin{array}{l}
		   \dot{u} = v\\
			\dot{v}  = -\eta v - \dfrac{\delta E}{\delta \bar{u}} .
	\end{array}
\end{equation}
Then, we can apply a symplectic explicit  method, such as symplectic Euler or St\"{o}rmer-Verlet~\cite{HairerLubichWanner2006}, which gives us an iterative map in the numerical approximations $(u_\nu, \ v_\nu), \ \nu=1, \: 2, \: 3 , \: ...  $ with a step in fictitious time $\Delta \tau_\nu$ and damping $\eta_\nu$.
The choice of parameters $\Delta \tau_\nu$ and $\eta_\nu$ can be chosen in order to optimize the performance of the numerical method, which generally is a non-trivial task.
However, for linear differential equations, such as the Schr\"{o}dinger equation, analytic results exist~\cite{Gulliksson_book_chapter_2019, Gulliksson2017}. 
For simplicity we will keep all parameters constant through the iterations, i.e., independent of the step~$\nu$.

The approach of finding the solution to Eq.~(\ref{eq:E_u}) by solving Eq.~(\ref{eq:dynamgV}) with a symplectic method has been named the {\em dynamical functional particle method} (DFPM)  \cite{GullikssonEdvardssonLind2013}.
We would like to emphasize that it is the combination of the second order damped dynamical system  together with an efficient (fast, stable, accurate)
symplectic solver that makes DFPM a very powerful method.
Even if the idea of solving minimization problems using dynamical systems with different damping strategies goes far back, see Refs.~\cite{MR016940319640101,IncertiValerioFranceso1979}, it has not been presented for the Schr\"{o}dinger equation with constraints with symplectic solvers for second order systems. 
According to our practical experience many common (non-symplectic) integration methods with optimal or non-optimal damping parameters give a reasonably fast convergence of Eq.~(\ref{eq:dynamgV}) to the stationary solutions.
So, unless the numerical performance is important, a variety of softwares for dynamical systems can be used in practical implementations.

Let us finally comment on one
closely related approach for solving Eq.~(\ref{eq:E_u}) that has been studied extensively~\cite{Schroeder_ajp_2017, SMYRLIS2004436} namely
\textit{the steepest descent method} 
\begin{equation}
\label{SD}
\dot{u} + \alpha \frac{\delta E}{\delta \bar{u}} = 0 ,\, \alpha>0,  \ \int |u|^2 d \mathbf{r} = 1.
\end{equation}
The method is called \textit{the imaginary time method} when applied for the SE with $\alpha=1$ (i.e., change $t \rightarrow - \mathrm{i}  \tau$ in Eq.~(\ref{eq:TDSE})).
It might seem that Eq.~(\ref{SD})
is better than Eq.~(\ref{eq:dynamgV}) since the exponential decrease towards the stationary solution in Eq.~(\ref{SD}) can be made arbitrary large  by choosing $\alpha$ large enough.
However, as proven strictly for linear problems~\cite{Gulliksson_book_chapter_2019}, and by numerical evidence for some nonlinear examples~\cite{SandinOgrenGulliksson2016}, going to a second order differential equation in a fictitious time is superior if one takes into account
the stability and accuracy of the numerical solver.
DFPM has been shown to have a remarkably faster
convergence to the stationary solution than any numerical method applied to Eq.~(\ref{SD}), see~\cite{Gulliksson_book_chapter_2019}.

It is the purpose of this article to explain this new method through a few qualitatively different examples for the stationary SE.
In addition it will be extended to a corresponding method to treat constraints.

\subsection{Damped oscillator approach for global constraints} 

DFPM is readily extended to more general constrained problems, e.g., normalized- and excited states of the SE in general settings.
Consider a convex minimization problem for $E(u)$ with smooth global constraint functionals $G_j(u) = 0$, i.e.,
\begin{equation}
\label{minVg}
 \min_{u} E(u), \: \textrm{s.t.} \: G_j(u)=0. 
\end{equation}
The actual choice of index $j$ is dependent on what is natural for different problem settings.
To give one example of a constraint $G_j(u) = 0$, the normalization constraint in Eq.~(\ref{eq:E_u}), can be written on this form, see Eq.~(\ref{app_A_G_1}).  

The problem in Eq.~(\ref{minVg}) has a unique solution $u^*$ if $\delta G_j / \delta \bar{u}$ is surjective at $u^*$, and thus fulfills the so-called Karush-Kuhn-Tucker conditions.
The corresponding dynamical system for constraints can be formulated  using an extended constrained energy functional (often called Lagrange function in mathematical literature)
$
   I(u,\mu_1, \mu_2, ...) = E(u) + \sum_j \mu_j G_j(u),
$
where $\mu_j$ are Lagrange multipliers.
The dynamical system for solving Eq.~(\ref{minVg}) is then given by 
\begin{equation}
\label{DFPMVg}
 \ddot{u} + \eta \dot{u} +  \frac{ \delta E }{\delta \bar{u}} +  \sum_j \mu_j\frac{ \delta G_j }{\delta \bar{u}} = 0,
\end{equation}
with $\mu_j (\tau)$ chosen such that $u(\tau)$ tends to the stationary solution $u^*$ when $\tau$ tends to infinity, see examples in the next section. 
For more details on existence and uniqueness of solutions to constrained problems see~\cite{McLachlanModinVerdierWilkins2014} and references therein.

In order to solve Eq.~(\ref{minVg}) one can choose the $\mu_j(\tau)$ such that $u(\tau)$ always remains on the constraints set, 
e.g., by projection methods, or as we will do here, to approach the constraints set (usually in a oscillatory manner) as $\tau$ increases. 
Projection is generally costly but there are important exceptions such as, e.g., eigenvalue problems with only normalization constraints.  

For our damped approach, we introduce an additional dynamical system, analogous to Eqs.~(\ref{eq:dharmosc}) and~(\ref{eq:dynamgV}), for a constraint $G_j$ as in Eq.~(\ref{minVg}) according to
\begin{equation}
\label{gdamp}
   \ddot{G}_j + \eta \dot{G}_j + k_j G_j = 0,\,  \eta>0, \, k_j>0. 
\end{equation}
Note that equations~(\ref{gdamp}) describes damped oscillators.
Then $G_j(u(\tau))$ tends to zero exponentially fast and the equations~(\ref{gdamp}) can be used to derive expressions of the Lagrange multipliers $\mu_j (\tau)$ for Eq.~(\ref{DFPMVg}), which for some problems are explicit, see further Secs.~\ref{Subsec_Normalization_constraint} and~\ref{Subsec_Normalization_constraint_and_one_orthogonalization_constraint}.

This method, based on Eq.~(\ref{gdamp}), was introduced in~\cite{Gulliksson2017} for solving matrix eigenvalue problems, where it was shown that $u(\tau)$ converges
asymptotically to the eigenvectors. 
It was also shown that the choice of $k_j$ in Eq.~(\ref{gdamp}) does not change the local convergence rate if
$k_j$ lie within a rather large range which is determined by the eigenvalues to the operator $\delta E / \delta \bar{u}$, see~\cite{Gulliksson2017} for details. 
In this article we always keep $k_j=k$ for all $j$ for simplicity, while using the freedom in different $k_j$ can further improve the numerical performance of the method.

Under these assumptions, the local convergence rate of the corresponding symplectic Euler with the optimal parameters~\cite{Gulliksson_book_chapter_2019} will be the same as for the projection approach.
However, while the two approaches  have the same local behavior it is not generally \textit{a priori} known which of these two methods is faster for a specific problem.
Further note that a general known disadvantage with projection is that large changes in the Lagrange multipliers require small timesteps.

\section{Two examples of constraints with explicit derivations of Lagrange multiplicators}

We include two examples of constraints here in detail for the readers who wish to understand the method. 
We begin by deriving the Lagrange multiplier for only one normalization constraint,
then we add only an orthogonalization constraint, i.e., what is needed to calculate the first excited state of the SE.

\subsection{Normalization constraint} \label{Subsec_Normalization_constraint}
Taking the first and second order derivatives of the normalization constraint
\begin{equation}
G_1 = 1 - \int |u|^2 d \mathbf{r} \equiv 1 - N(\tau) = 0, \label{app_A_G_1}
\end{equation}
with respect to $\tau$ gives
\begin{equation}
\dot{G}_1= - \int \left( \dot{ \bar{u} } u +  \bar{u} \dot{u} \right) d \mathbf{r}  , \ \ddot{G}_1 =  - \int \left( \ddot{ \bar{u} } u  +  2 \dot{ \bar{u} } \dot{u}  +   \bar{u} \ddot{u} \right) d \mathbf{r}. \label{app_A_ddot_and_dot_G_1}
\end{equation}
Inserting the expressions from Eq.~(\ref{app_A_ddot_and_dot_G_1}) into the left hand side of the general differential equation for constraints~(\ref{gdamp}), then gives after simplifications  
\begin{equation}
\ddot{G}_1 + \eta \dot{G}_1 =  - \int \left( \left\{ \ddot{ \bar{u} } + \eta \dot{ \bar{u} } \right\} u +  \bar{u}  \left\{ \ddot{ u } + \eta \dot{ u } \right\}  + 2 |\dot{u}|^2 \right) d \mathbf{r}.    \label{app_A_dyn_eq_G_1}
\end{equation}
If we now use the DFPM equation~(\ref{DFPMVg}) with $ \delta E / \delta \bar{u} + \mu_1 \delta G_1 / \delta \bar{u}= \hat{H}u-\mu_1 u$ for the stationary SE.
We can then, when $ \ddot{ u } + \eta \dot{ u } \rightarrow 0 $, identify the limit of the Lagrange multiplier being equal to the energy, $\lim_{\tau \rightarrow \infty}\mu_1 \left( \tau \right) =E$, compare with Eq.~(\ref{eq:SE}). 
Inserting $- \hat{H}u + \mu_1 u$ into the curly brackets of Eq.~(\ref{app_A_dyn_eq_G_1}) gives after simplifications
\begin{equation}
\ddot{G}_1 + \eta \dot{G}_1 =  2E - 2\mu_1 N -  2 \int |\dot{u}|^2  d \mathbf{r} = - k_1 \left( 1 - N \right),    \label{app_A_simplified_dyn_eq_G_1}
\end{equation}
with $E ( \tau ) \equiv \int \bar{u} \hat{H} u  \, d \mathbf{r}$.
Finally we can solve for the Lagrange multiplier $\mu_1 \left( \tau \right)$
\begin{equation}
\mu_1=  \frac{ E + k_1 \left( 1 - N \right)/2 - \int |\dot{u}|^2  d \mathbf{r}}{N}.   \label{app_A_mu_1}
\end{equation}
We see in Eq.~(\ref{app_A_mu_1}) that $\mu_1 \rightarrow E$, since $|\dot{u}| \rightarrow 0, \ N \rightarrow 1$ as $\tau \rightarrow \infty$.

\subsection{Normalization constraint and one orthogonalization constraint} \label{Subsec_Normalization_constraint_and_one_orthogonalization_constraint} 

Introducing, in addition to $G_1$ above, the following orthogonalization constraint
\begin{equation}
G_0= \int \bar{u} u_0 d \mathbf{r} = 0, \label{app_A_G_2}
\end{equation}
means that the solution $u$, should be orthogonal to a known normalized function $u_0$. 
This $u_0$ can be defined analytically, which can be helpful while testing software~\cite{program}, but more often $u_0$ is a numerically obtained approximation. 
For example in the case of a convex 1D problem, $u_0$ is the solution with the lowest eigenvalue $E$ (groundstate), and $u$ is the solution with the second lowest eigenvalue $E$ (first excited state). As seen in the 2D example of Sec.~\ref{sec-2DHO}, this situation can be more complicated in higher dimensions where eigenvalues can be degenerate, meaning that several different solutions can have the same eigenvalue.

Taking the first and second order derivatives with respect to $\tau$ of the orthogonalization constraint in Eq.~(\ref{app_A_G_2}) gives 
\begin{equation}
\dot{G}_0= \int  \dot{ \bar{u} } u_0 d \mathbf{r}  , \ \ddot{G}_0=  \int \ddot{ \bar{u} } u_0 d \mathbf{r}. \label{app_A_ddot_and_dot_G_2}
\end{equation}
Inserting the expressions from Eqs.~(\ref{app_A_G_2}) and~(\ref{app_A_ddot_and_dot_G_2}) into the general differential equation for constraints~(\ref{gdamp}), then gives with simplifications
\begin{equation}
\ddot{G}_0 + \eta \dot{G}_0 =  \int \left\{ \ddot{ \bar{u} } + \eta \dot{ \bar{u} } \right\} u_0 d \mathbf{r} = \int \left\{ -\hat{H} \bar{u}  + \mu_1 \bar{u} \right\} u_0 d \mathbf{r}     - \mu_0 = - k_0 G_0 .    \label{app_A_dyn_eq_G_2}
\end{equation}
In comparison to Eq.~(\ref{app_A_simplified_dyn_eq_G_1}) there is now an additional term ($j=0$) in Eq.~(\ref{DFPMVg}).
The corresponding coupled equation for $G_1$ is
\begin{equation}
\ddot{G}_1 + \eta \dot{G}_1 =  2E - 2\mu_1 N  +  \int \left(  \mu_0 \bar{u}_0 u  + \mu_0 \bar{u} u_0 -  2 |\dot{u}|^2  \right) d \mathbf{r} = - k_1 G_1.   \label{app_A_simplified_dyn_eq_G_1_for_G_2}
\end{equation}
Finally we write Eqs.~(\ref{app_A_dyn_eq_G_2}) and~(\ref{app_A_simplified_dyn_eq_G_1_for_G_2}) for the two coupled Lagrange multipliers $\mu_0 \left( \tau \right)$ and $\mu_1\left( \tau \right)$ as a linear system 
\begin{equation}
\begin{bmatrix}
 1  & -G_0 \\
	-\operatorname{Re} \left( G_0 \right) & 1-G_1   \\ 
\end{bmatrix}
\begin{bmatrix}
	\mu_0 \\
	\mu_1 \\
\end{bmatrix}
=
\begin{bmatrix}
	k_0 G_0 - \int u_0 \hat{H} \bar{u} d \mathbf{r}  \\
	E  - \int |\dot{u}|^2  d \mathbf{r} +  k_1 G_1 /2 \\
\end{bmatrix}
\equiv
\begin{bmatrix}
	y_1 \\
	y_2 \\
\end{bmatrix}
,  \label{app_A_mu_1_and_mu_2}
\end{equation}
where $\operatorname{Re} \left( G_0 \right) =  \int \left( \bar{u}_0 u + \bar{u} u_0 \right) d \mathbf{r} /2 $.
We can check the limits for the Lagrange multipliers from Eq.~(\ref{app_A_mu_1_and_mu_2}),  
i.e., $\mu_0 \rightarrow 0$ and $\mu_1 \rightarrow E$, since $\dot{u} \rightarrow 0, \: G_{0} \rightarrow 0, \: G_{1} \rightarrow 0$ as $\tau \rightarrow \infty$.

Using Cramer's rule on the linear system of Eq.~(\ref{app_A_mu_1_and_mu_2}) give the explicit expressions
\begin{equation}
\mu_0 =  \frac{ (1-G_1) y_1 + G_0 y_2 }{1-G_1 - G_0 \operatorname{Re} \left( G_0 \right)}, \ \: \mu_1 = \frac{ y_2 +  \operatorname{Re} \left( G_0 \right) y_1 }{1-G_1 - G_0 \operatorname{Re} \left( G_0 \right)}
, \label{explicit_general_LM_for_first_excited_state}
\end{equation}
that can be used to calculate the first excited state in various problems. 

In~\ref{Sec_appendix_A}, we show how an arbitrary number of orthogonalization constraints are treated.
We stress that the obtained result from~\ref{Sec_appendix_A} can be used for 
Schr\"{o}dinger equations in any dimension with any potential, as illustrated in Sections~\ref{sec-Hydrogen} and~\ref{sec-2DHO}.

\section{Numerical examples}

In this section we show numerical results using the symplectic Euler method~\cite{HairerLubichWanner2006} for three examples and compare the presented DFPM method against analytic formulas.
The first example is a linear equation in one radial variable for the hydrogen atom.
The second is a harmonic oscillator in two variables (2D), which gives degenerated solutions.
Finally, an example of a nonlinear Schr\"{o}dinger equation is given.

\subsection{The radial equation for the hydrogen atom}  \label{sec-Hydrogen} 
The function $u$, is in this example the radial part of the three-dimensional spatial wavefunction from Eq.~(\ref{eq:TDSE}), but multiplied with the radius $r= |\mathbf{r}|$, i.e., $\tilde{u} (\mathbf{r}) = \tilde{u} (r,\theta,\phi) =u(r)/r \, Y_l^m(\theta,\phi)$, where $Y_l^m$ are the spherical harmonics.
We write the dimensionless (i.e., with $\hbar=M_e=a_0=1$) radial SE of the hydrogen atom as
\begin{equation}
\hat{H}_{(r)} u \equiv  -\frac{1}{2} \frac{d^2 u}{d r^2} + \left( \frac{l \left( l + 1 \right)}{2 r^2}  -  \frac{1}{r} \right) u = E u, \ 4 \pi \int_0^\infty |u|^{2} d r =  1, \label{eq:Hamiltonian_Hydrogen}
\end{equation}
where the expression within the large paranthesis in Eq.~(\ref{eq:Hamiltonian_Hydrogen}) is referred to as the effective radial potential.
The factor $4 \pi$ in the normalization constraint comes from the radial symmetry in 3D. 
For comparison, the energies only depend on a single quantum number $n=1,2,3,...$ and are given by~\cite{ref_solutions_to_hydrogen_and_2d_ho}
\begin{equation}
E_n = -\frac{M_e e^4}{2 \left( 4 \pi \varepsilon_0 \right)^2 \hbar^2} \frac{1}{n^2}  =  - \frac{\hbar^2}{2 M_e a_0^2} \frac{1}{n^2} =  - \frac{1}{2} \frac{1}{n^2}, \label{eq:Energies_Hydrogen}
\end{equation}
where $a_0 = 4\pi \varepsilon_0 \hbar^2/(M_e e^2)$ is a length scale called the Bohr radius.
The corresponding radial wavefunctions depend on the two quantum numbers $n=1,2,3,...$ and $l=0,1,2,...,n-1$ and are given by~\cite{ref_solutions_to_hydrogen_and_2d_ho}
\begin{equation}
u_{n,l}(r)/r = \sqrt{ \frac{1}{\pi a_0^3 n^4} \frac{ \left( n-l-1 \right)!}{\left(n+l \right)!} } \left( \frac{2r}{n a_0} \right)^l L_{n-l-1}^{2l + 1} \left( \frac{2r}{n a_0} \right) \exp \left(-\frac{r}{n a_0}\right) , \label{eq:Solutions_Hydrogen}
\end{equation}
where $L$ denotes the generalized Laguerre polynomials~\cite{ref_special_functions}, 
defined in agreement with the \textsc{Matlab} command \texttt{laguerreL(n-l-1,2*l+1,2*r/n)}, although different normalization factors can be found in the physics literature. 
\begin{figure}
\hspace{24mm} 
\includegraphics[scale=0.6]{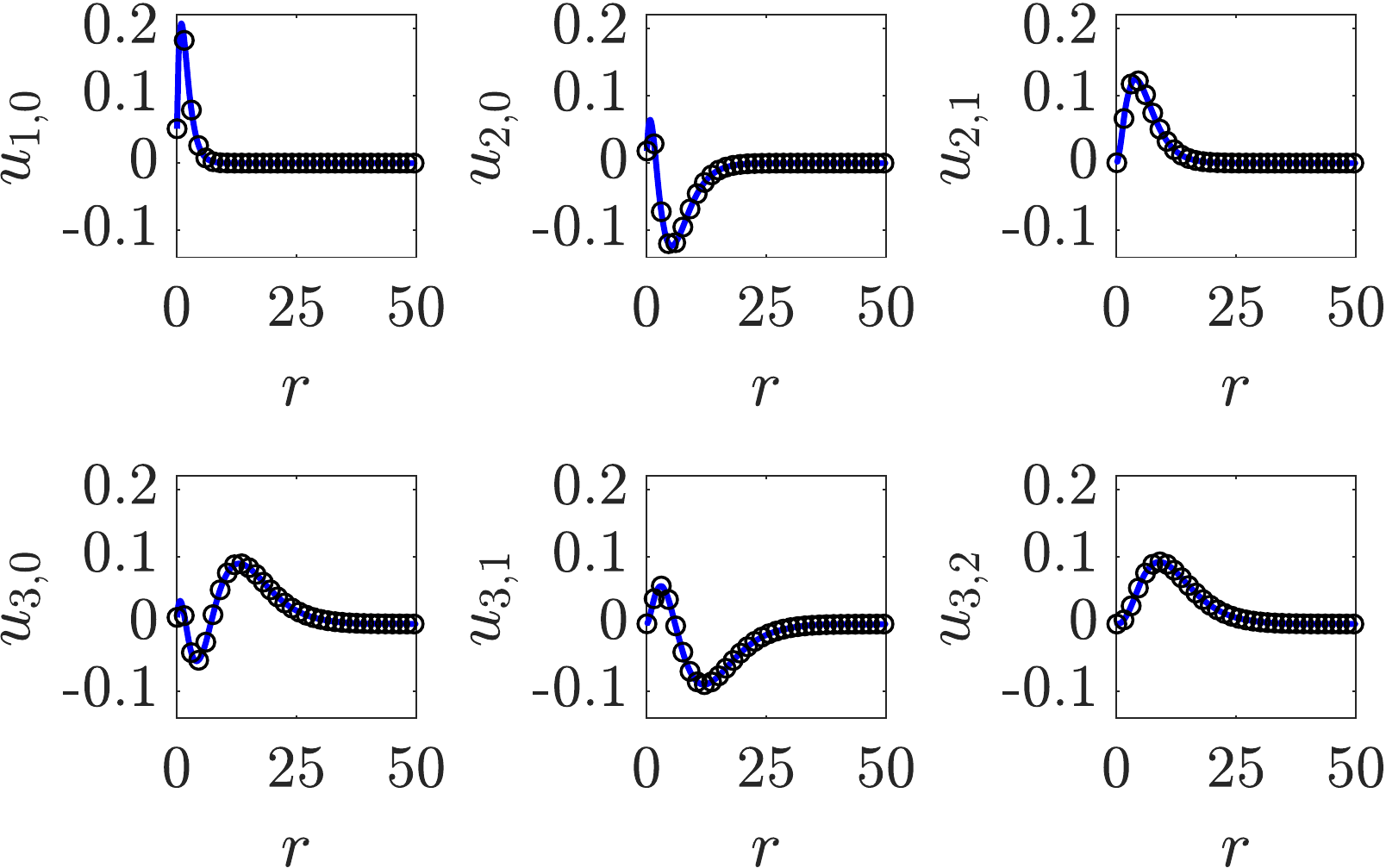} 
\caption{The six numerical solutions of Eq.~(\ref{eq:SE_Hydrogen_DFPM}) with the lowest energies $E$.
Solid (blue) curves are numerical results, while (black) circles show the result of Eq.~(\ref{eq:Solutions_Hydrogen}).
We used an equidistant grid $10^{-6} < r_j < 10^{2}, \: j=1, \: ..., \: 10^3$, and $\tau_{\max}$ large enough such that $| \ddot{u} + \eta \dot{u}| < 10^{-6}$ in Eq.~(\ref{DFPMVg}).
DFPM parameters used were $\eta=0.5$, $k= 4$, and $\Delta \tau= 0.1$, which are of the same order of magnitude as the optimal predicted values for linear systems~\cite{Gulliksson_book_chapter_2019}.
The (real) initial condition $u(0)$ was in this example chosen randomly.
We note that the sign of the final wavefunction depends on the initial condition used. 
}\label{Fig_2019_wavefunctions_hydrogen}
\end{figure}

The effective radial potential in Eq.~(\ref{eq:Hamiltonian_Hydrogen}) depends on the quantum number $l$, as do the solutions in Eq.~(\ref{eq:Solutions_Hydrogen}), so there is not any degeneracy when solving Eq.~(\ref{eq:Hamiltonian_Hydrogen}) numerically.
In other words, the solution of Eq.~(\ref{eq:Hamiltonian_Hydrogen}) is unique for this radial SE.
However, all states with the same quantum number $n$ have the same energy $E$, as is clear from Eq.~(\ref{eq:Energies_Hydrogen}),
and together with the degeneracy ($2m+1$) for the spherical harmonics $Y_l^m$, the three-dimensional wavefunction for Hydrogen have a $n^2$ degeneracy.

The solutions $u_{n,l^*}$ with $n=1,2,...,n^*-1$ should be orthogonal to the unknown $u_{n^*,l^*}$.
Since $u_{n, l^*},\: \ n=1,\:2,\:...,n^*-1$ are needed in order to obtain $u_{n^*, l^*}$,
we solve Eq.~(\ref{DFPMVg}) in consecutive order. 
More specifically, we start with the normalization constraint, see Sec.~\ref{Subsec_Normalization_constraint}, to obtain $u_{1, l^*}$, then add an orthogonality constraint, see Sec.~\ref{Subsec_Normalization_constraint_and_one_orthogonalization_constraint}, to obtain $u_{2, l^*}$, and generally several orthogonality constraints, see~\ref{Sec_appendix_A}, to obtain the solutions $u_{n^*, l^*}$ with $n^*>2$.

We can write the $n^*$ constraints compactly using the Kronecker delta as $G_{n} = 4 \pi \int_0^\infty \bar{{u}}_{n^*,l^*} u_{n,l^*} dr - \delta_{n^*,n} = 0,\: n=1,\:2,\:..., \: n^*$. 
Using Eq.~(\ref{eq:Hamiltonian_Hydrogen}) we formulate the dynamical system, different for each value of $l^*$, that is Eq.~(\ref{DFPMVg}) applied to this radial SE is
\begin{equation}
\label{eq:SE_Hydrogen_DFPM}
\ddot{u}_{n^*,l^*} +\eta \dot{u}_{n^*,l^*} +  \hat{H}_{(r)}(l^*) u_{n^*,l^*} + \sum_{n=1}^{n^*} \mu_n \frac{\delta G_n}{\delta \bar{u}} = 0.
\end{equation}
The two Lagrange multipliers needed in the sum above to calculate the first excited state, i.e. for $n^*=2$, can be obtained from Eq.~(\ref{explicit_general_LM_for_first_excited_state}).
For the case with several multipliers ($n^* > 2$) in Eq.~(\ref{eq:SE_Hydrogen_DFPM}), they can conveniently be obtained from, e.g., numerical solutions of Eq.~(\ref{app_A_lin_sys}). 

\begin{figure}
\hspace{24mm} 
\includegraphics[scale=0.55]{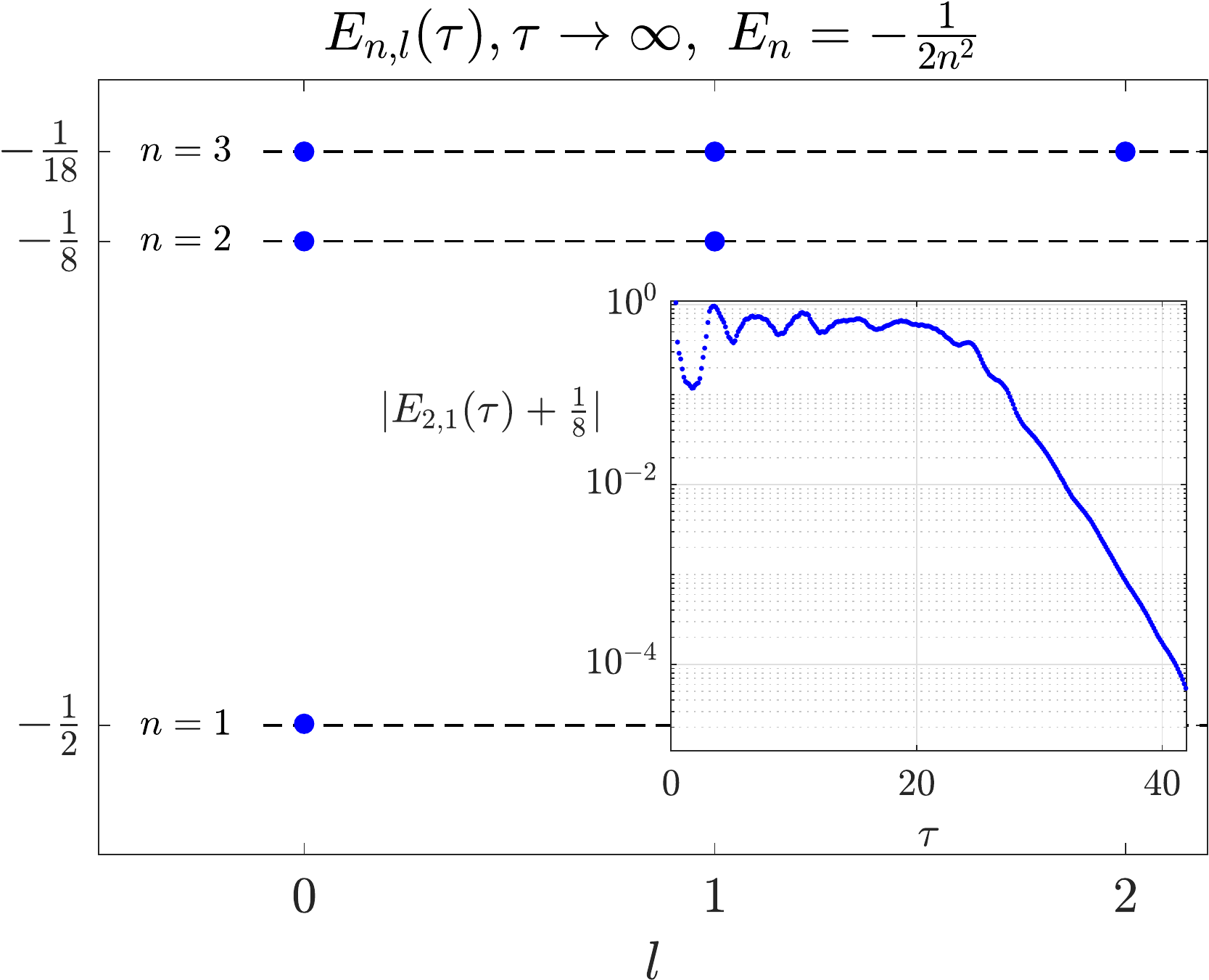} 
\caption{The six lowest energies $E$ for the hydrogen atom.
Dots (blue) are the numerically calculated energies from Eq.~(\ref{eq:SE_Hydrogen_DFPM}).
The dashed horizontal (black) lines correspond to the values from Eq.~(\ref{eq:Energies_Hydrogen}).
The inset figure shows the convergence dynamics of the energy $E_{2,1}$ as function of the fictitious time.
The parameters used are the same as the one in the caption of Fig.~\ref{Fig_2019_wavefunctions_hydrogen}.
}
 \label{Fig_2019_energies_hydrogen}
\end{figure}

The six stationary numerical solutions to Eq.~(\ref{eq:SE_Hydrogen_DFPM}) with lowest energies are plotted in Fig.~\ref{Fig_2019_wavefunctions_hydrogen}, 
while the corresponding energies are illustrated in Fig.~\ref{Fig_2019_energies_hydrogen}. 

\subsection{Two-dimensional harmonic oscillator} \label{sec-2DHO} 
In this example we calculate the well known wave functions $u(x,y)$ and energies $E$
to the dimensionless (i.e., with $\hbar=M=\omega=1$) Schr\"{o}dinger equation with an isotropic two-dimensional harmonic potential on a 2D Cartesian grid
\begin{equation}
\hat{H} u \equiv -\frac{1}{2} \left( \frac{\partial^2 }{\partial x^2} + \frac{\partial^2 }{\partial y^2} \right)u +\frac{1}{2} \left( x^{2} + y^{2} \right) u=E u, \ \int\limits_{\mathbb{R}^2} |u|^{2}d \mathbf{r}=  1.\label{eq:2D_HO_SE}
\end{equation}
That is, for the groundstate we solve the following optimization problem
\begin{equation}
 \min\limits_u     \int\limits_{\mathbb{R}^2}\bar{{u}} \hat{H} u \, d\mathbf{r}  ,\:  \textrm{s.t.}  \int\limits_{\mathbb{R}^2} |u|^{2} \, d\mathbf{r} =  1.
\end{equation}

\begin{figure}
\hspace{24mm} 
\includegraphics[scale=1.0]{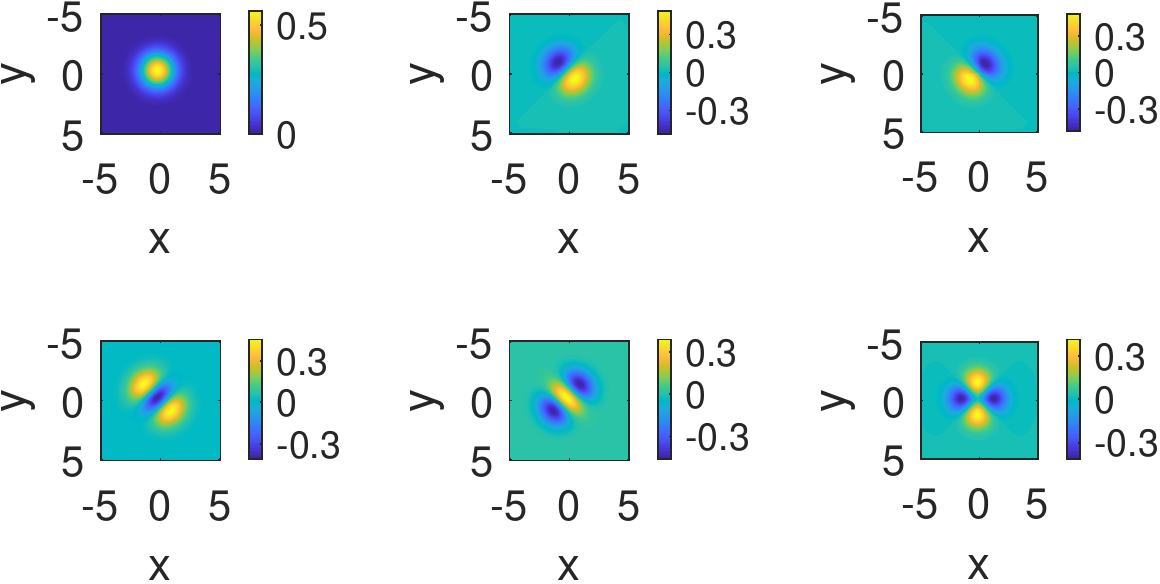} 
\caption{The six numerical wavefunctions of Eq.~(\ref{eq:SE_HO_DFPM_12}), with the lowest energies $E$.
We used $-5 \leq x, \: y \leq 5$ and $\Delta x= \Delta y= 1/12$ for the discretization, which is enough to obtain all six solutions with correct energies within 3 significant digits. 
The parameters were $\eta=1.5, \ k=0.5$ and $\Delta \tau =0.05$, which is in the same order of magnitude as predicted to be optimal~\cite{Gulliksson_book_chapter_2019}. 
The initial wavefunction (for all six subfigures here) was a translated and scaled (unnormalized) Gaussian $u( \tau=0 ) = 1.2/\sqrt{ \pi } \exp(-( (x-1.2)^2 + (y-1.2)^2)/2 )$ with $E( \tau=0 ) \simeq 2.5$ (see the left ring in Fig.~\ref{Fig_se_ho_energies_and_orthonormalization}).
The six numerical wavefunctions, from the upper left subfigure to the bottom right subfigure, corresponds to $u_{(n_x,n_y)}$ from Eq.~(\ref{eq:SE_HO_analytic}) in the order
$(n_x,n_y)=(0,0), \ (1,0),\: (0,1); \ (2,0),\: (0,2) ,\: (1,1)$. 
However, note that the orientation and phase (sign) of the final wavefunction depends on the initial condition used. 
}
 \label{Fig_se_ho_eigenstates}
\end{figure}
Let $u_{s}$ denote the $s$'th eigenstate and  $E_{s} = \int\limits_{\mathbb{R}^2}\bar{{u}}_{s} \hat{H} u_{s} d\mathbf{r}$ the corresponding eigenvalue.
To obtain $u_{s^*}$ for $s^*>1$ we use, in addition to Eq.~(\ref{eq:2D_HO_SE}), the $s^*-1$ orthogonality constraints
\begin{equation}
\int\limits_{\mathbb{R}^2} \bar{u}_{s^*}  u_1 d\mathbf{r}=0,\:  \int\limits_{\mathbb{R}^2} \bar{u}_{s^*}  u_2 d\mathbf{r}=0,\: ...,\:\int\limits_{\mathbb{R}^2} \bar{u}_{s^*}  u_{ s^*-1 } d\mathbf{r}=0. \label{eq:orthogonality_conditions}
\end{equation}
Using Eqs.~(\ref{eq:2D_HO_SE}) and (\ref{eq:orthogonality_conditions})
we can from Eq.~(\ref{DFPMVg}) formulate the corresponding dynamical system
with the $s^*$ constraints $G_{s}=\int\limits_{\mathbb{R}^2} \bar{u}_{ s^* } u_{ s } d\mathbf{r} - \delta_{s^*, s}=0,\: s=1,\:2,\:...,s^*$, as 
\begin{equation}
\label{eq:SE_HO_DFPM_12}
\ddot{u}_{ s^* } + \eta \dot{u}_{ s^* } +  \hat{H} u_{ s^* } +   \sum_{s=1}^{s^*} \mu_s \frac{\delta G_s}{\delta \bar{u}}= 0. 
\end{equation}
Since one needs access to $u_{ s },\: s=1,\:2,\:...,s^*-1$,
we solve Eq.~(\ref{eq:SE_HO_DFPM_12}) in consecutive order. 
The two Lagrange multipliers needed in the sum above to calculate the first excited state, i.e. for $s^*=2$, are given by Eq.~(\ref{explicit_general_LM_for_first_excited_state}).
For the case with several multipliers ($s^* > 2$) in Eq.~(\ref{eq:SE_HO_DFPM_12}), see~\ref{Sec_appendix_A}, they can conveniently be obtained from e.g. numerical solutions of Eq.~(\ref{app_A_lin_sys}). 
We show the six first numerical solutions to Eq.~(\ref{eq:SE_HO_DFPM_12}) in Fig.~\ref{Fig_se_ho_eigenstates}.

For comparisons we note that the equation (\ref{eq:2D_HO_SE}) has the explicit solutions~\cite{ref_solutions_to_hydrogen_and_2d_ho} 
\begin{equation}
u_{(n_x,n_y)} \left( x,y \right) =   \frac{1} {\sqrt{2^{\left( n_x+n_y \right)} n_x! n_y! \pi} } \mathcal{H}_{n_x}\left(x\right) \mathcal{H}_{n_y}\left(y\right) \exp\left(-\frac{x^{2}+y^{2}}{2}\right),\label{eq:SE_HO_analytic}
\end{equation}
where $\mathcal{H}$ denote the Hermite polynomials~\cite{ref_special_functions}, and the two quantum numbers can take the values $n_x,\: n_y = 0,\: 1,\:2,\:3,\:...$ .

In contrast to the radial SE for the hydrogen atom, there is no dependence on any of the quantum numbers $n_x, \: n_y$ in the SE~(\ref{eq:2D_HO_SE}), and different solutions $u_{(n_x,n_y)}$ can give degenerate energies $E_{(n_x,n_y)}= n_x+n_y+1$ as long as $n_x + n_y$ is $\textrm{constant}$. 

\begin{figure}
\hspace{24mm} 
\includegraphics[scale=0.4]{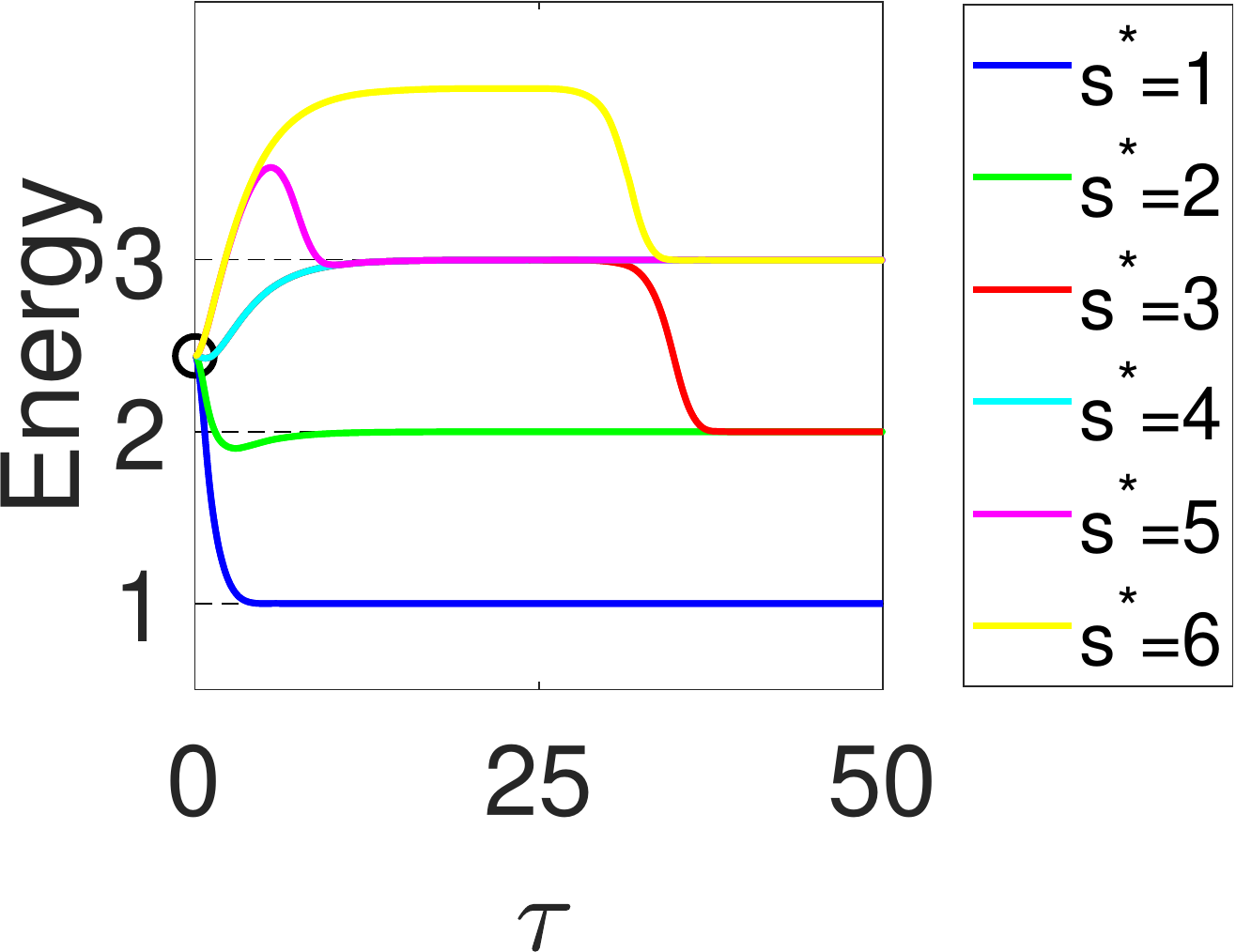}~~~\includegraphics[scale=0.4]{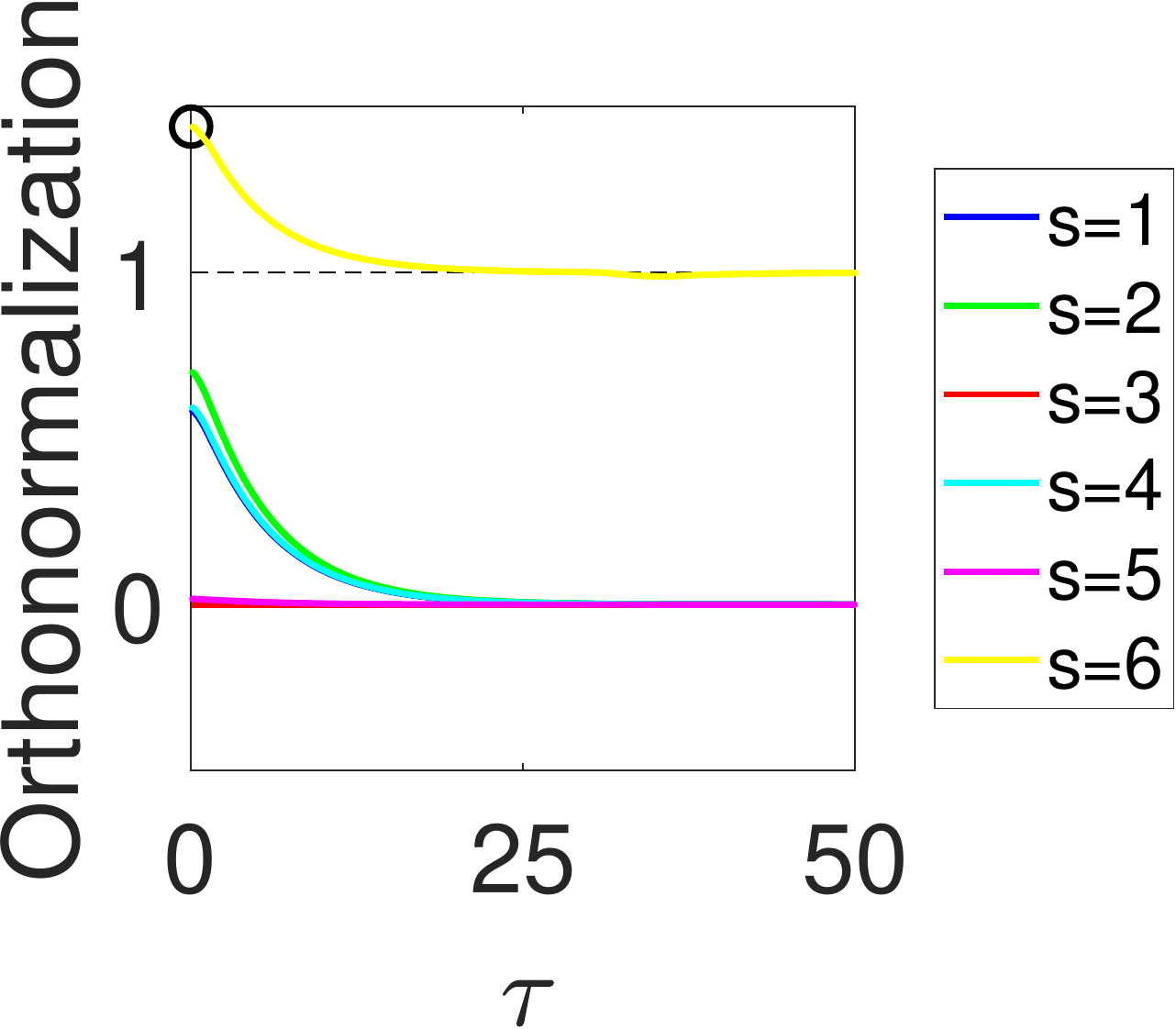}
\caption{ Left figure: Convergence of the energies for the six solutions seen in Fig.~\ref{Fig_se_ho_eigenstates}.
Dashed horizontal lines correspond to the exact energies $E_{(n_x,n_y)}= n_x+n_y+1$. 
Right figure: Illustration of the convergence of the calculation for solution number six (i.e. $s^*=6$ corresponding to $n_x=n_y=1$).
The most upper curve show the normalization $\int |u|^2 d\mathbf{r}$ (the ring shows the value $1.2^2=1.44$, see the initial condition in the caption of Fig.~\ref{Fig_se_ho_eigenstates}), and the lower curves shows the five orthogonality constraints $\int \bar{u}  u_{ s } d\mathbf{r}, \: s=1,2,3,4,5$.
} \label{Fig_se_ho_energies_and_orthonormalization}
\end{figure}
In the left plot of Fig.~\ref{Fig_se_ho_energies_and_orthonormalization} we show the numerical convergence for the energies.
In the right plot of Fig.~\ref{Fig_se_ho_energies_and_orthonormalization} we show the numerical convergence for the constraints.

\subsection{The nonlinear Schr\"{o}dinger equation under rotation}

The nonlinear Schr\"{o}dinger equation (NLSE) is commonly used to model many interacting bosonic particles via a mean-field approximation~\cite{PethickSmithBEC}. 
We have developed a DFPM formulation with damped constraints for a dimensionless
nonlinear Schr\"{o}dinger equation in $u=u(x)$ on a ring geometry $-\pi \leq x \leq \pi$
($R=\hbar= 2 M = 1$) using periodic boundary conditions $u(-\pi)=u(\pi)$.

The aim is to minimize the total energy
\begin{equation}
E(u) = \int\limits_{-\pi}^{\pi}  \left|\frac{ \partial u }{\partial x}\right|^{2} + \pi \gamma \left|u\right|^{4} dx,\label{eq:NLSE_total_energy}
\end{equation}
with $\gamma$ a parameter for the nonlinear term,
subject to one constraint for normalization, and one constraint for the angular momentum being~$\ell_0$
\begin{equation}
G_{1}= 1 - \int\limits_{-\pi}^{\pi}\left|u\right|^{2}dx = 0,\ G_{2}= \ell_0 + \mathrm{i}
\int\limits_{-\pi}^{\pi}\bar{{u}}   \frac{ \partial u}{\partial x}   dx =0. \label{constraints_for_NLSE}
\end{equation}
We note that this problem can be solved analytically and refer to Appendix~B of Ref.~\cite{SandinOgrenGulliksson2016} for the details of the solutions.
In earlier work we implemented DFPM numerically for this problem with a modified RATTLE method~\cite{SandinOgrenGulliksson2016}, in which
we solved for the two Lagrange multipliers corresponding to Eq.~(\ref{constraints_for_NLSE}) numerically in each timestep. 
There it was demonstrated that DFPM outperformed another commonly used method that is first order in time~\cite{SandinOgrenGulliksson2016}.
In this article we instead couple the minimization of Eq.~(\ref{eq:NLSE_total_energy}) to Eq.~(\ref{constraints_for_NLSE}), 
via the dynamical equations~(\ref{gdamp}) for the constraints and get the following realization of Eq.~(\ref{DFPMVg}) 
\begin{equation}
\ddot{u}+\eta \dot{u}  +  \frac{\delta E}{\delta \bar{u} } + \mu \frac{\delta G_1}{\delta \bar{u} } + \Omega \frac{\delta G_2}{\delta \bar{u} }   =  \ddot{u}+\eta \dot{u}   - \frac{ \partial^2 u}{\partial x^2} + 2\pi\gamma\left|u\right|^{2}u - \mu u + \mathrm{i}\Omega \frac{ \partial u}{\partial x} = 0 , \label{DE_for_NLSE_with_constraints_for_NLSE}
\end{equation}
with the two Lagrange multipliers from Eq.~(\ref{eq_app_B_linear_system_NLSE})
\begin{equation}
\mu =  \frac{b_1 \langle \hat{\ell}^2 \rangle - b_2 \ell}{N \langle \hat{\ell}^2 \rangle - \ell^2}, \ \: \Omega =  \frac{b_2 N - b_1 \ell}{N \langle \hat{\ell}^2 \rangle - \ell^2}. \label{LM_for_NLSE_with_constraints_for_NLSE}
\end{equation}
The Lagrange multipliers in Eqs.~(\ref{DE_for_NLSE_with_constraints_for_NLSE}) and~(\ref{LM_for_NLSE_with_constraints_for_NLSE}) represents physical properties.
Hence, $\mu$ is the so called chemical potential, which is not equal to the energy $E$ for the NLSE, and $\Omega$ is the angular velocity for the rotation. 
The quantities $N, \:  \langle \hat{\ell}^2 \rangle , \: \ell , \: b_1 , \: b_2$ in Eq.~(\ref{LM_for_NLSE_with_constraints_for_NLSE}), which depend on the fictitious time $\tau$, are defined in~\ref{Sec_appendix_B}. 

\begin{figure}
\hspace{24mm} 
\includegraphics[scale=0.55]{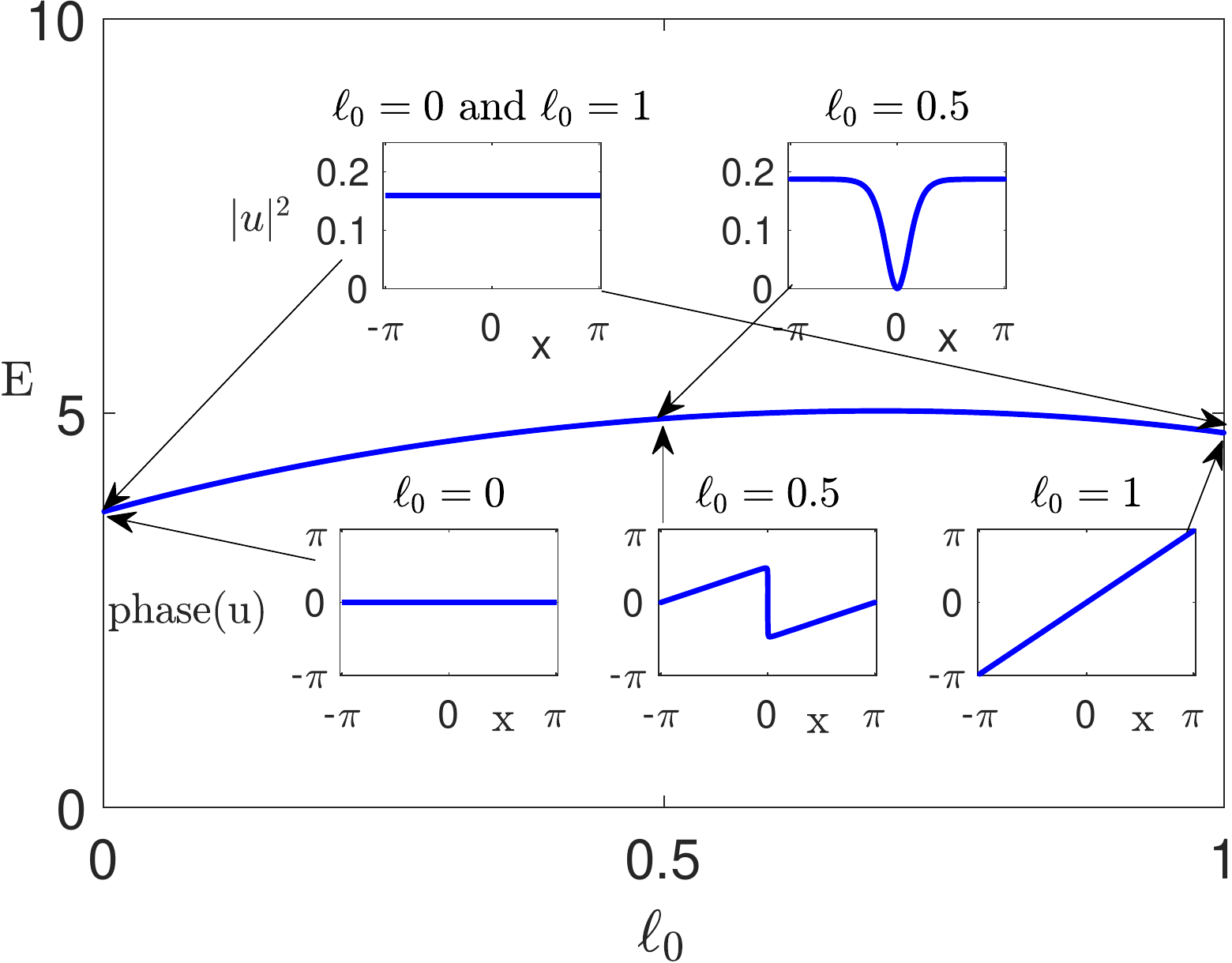}
\caption{Yrast curve~\cite{SandinOgrenGulliksson2016}, i.e. energy vs momentum, with some examples of the density
and the phase for the wavefunction $u$ for the constant of nonlinearity being $\gamma=7.5$. 
Optimal numerical parameters in Eq.~(\ref{DE_for_NLSE_with_constraints_for_NLSE}) are not trivially given in the nonlinear case, and we used $\eta = k/2 =1$ and $\Delta \tau = 0.015$. The spatial equidistant discretization consisted of 400 points}
\label{Fig_Yrast_curve}
\end{figure}

In Fig.~\ref{Fig_Yrast_curve} we have plotted the resulting so called
Yrast curve (main figure) with the density and phase of the corresponding
complex wave function $u$ for the particularly interesting points
$\ell_0=0,\:0.5,\:1$ (inset figures).
At integer values of $\ell_0$ ($0,\:1$ in this example), $u$ is a plane-wave $u=\exp\left(i\ell_0 x\right)/\sqrt{2\pi}$.
At half-integer values (e.g. $\ell_0=0.5$), $u$ corresponds to a dark
solitary wave that circulates in the ring~\cite{Jackson_et_al_EPL_2011}, see the right-upper- and mid-lower-inset figures.

\section{Conclusions}

We have introduced the \textit{dynamical functional particle method} (DFPM) with normalization and several orthogonalization constraints for the linear Schr\"{o}dinger equation. 
Numerical results are presented for the wavefunctions and energies of the radial part of the hydrogen atom, and for the 2D harmonic oscillator. 
Furthermore, DFPM was formulated with constraints for rotational states to the nonlinear Schr\"{o}dinger equation and then solved numerically.
 
We believe this presentation of DFPM may be helpful for students and researchers who want to solve globally constrained equations in general. 
More specifically, it can be used for numerically solving different kinds of Schr\"{o}dinger equations attaining (degenerated) excited states and energies. 

Obviously, DFPM will not outperform all existing methods for solving the large variety of different SE. However, as partly discussed in the introduction, we here mention four general advantages that we believe are of importance: 
\textit{1.}~The method has a general formulation and can solve many different kinds of problems.	
\textit{2.}~Among methods for solving equations with ordinary differential equations DFPM seems to be the best.
\textit{3.}~We have here specifically shown that DFPM is relevant and competitive for several important problems in quantum mechanics.
\textit{4.}~The method is simple to implement: Discretize in space then solve  the second order damped dynamical system with a stable method preferably a symplectic method, see the codes in the supplement material~\cite{program}.

\ack 

We acknowledge valuable comments from Patrik Sandin, and from two anonymous referees. 
We also thank the three ``French musketeers'' Julien R\'{e}gnier, Nico Gaudy, and Alexandre Clercq for valuable discussions about DFPM during their internships at \"{O}rebro University.

\appendix

\section{Normalization constraint and several orthogonalization constraints} \label{Sec_appendix_A}

In the numerical examples in Sections~\ref{sec-Hydrogen} and~\ref{sec-2DHO} both a normalization constraint and several orthogonalization constraints are treated simultaneously.
We here sketch how an arbitrary number of orthogonalization constraints is treated.
The result will hold for any potential $V(\mathbf{r})$ in any dimension.

We generalize Eqs.~(\ref{app_A_G_1}) and~(\ref{app_A_G_2}) to a vector containing $w$ orthogonalization constraints and one normalization constraint
\begin{equation}
\vec{G}= 
\begin{bmatrix}
	\int \bar{u} u_0 d \mathbf{r} \\
	\int \bar{u} u_1 d \mathbf{r} \\
	\vdots \\
	\int \bar{u} u_{w-1} d \mathbf{r} \\
	1 - \int \bar{u} u d \mathbf{r}  \\
\end{bmatrix}
= \vec{0}.
\end{equation} 
Hence with
\begin{equation}
\dot{\vec{G}}= 
\begin{bmatrix}
	\int \dot{\bar{u}} u_0 d \mathbf{r} \\
	\int \dot{\bar{u}} u_1 d \mathbf{r} \\
	\vdots \\
	\int \dot{\bar{u}} u_{w-1} d \mathbf{r} \\
	- \int \dot{\bar{u}} u +  \bar{u} \dot{u}  d \mathbf{r}  \\
\end{bmatrix}
,
\ddot{\vec{G}}= 
\begin{bmatrix}
	\int \ddot{\bar{u}} u_0 d \mathbf{r} \\
	\int \ddot{\bar{u}} u_1 d \mathbf{r} \\
	\vdots \\
	\int \ddot{\bar{u}} u_{w-1} d \mathbf{r} \\
	-\int \ddot{\bar{u}} u  +  \bar{u} \ddot{u}  +   2\dot{\bar{u}} \dot{u} d \mathbf{r}  \\
\end{bmatrix}
,
\end{equation} 
we have from Eq.~(\ref{gdamp})
\begin{equation}
\ddot{\vec{G}} + \eta \dot{\vec{G}}= 
\begin{bmatrix}
	\int \left( \ddot{\bar{u}} + \eta \dot{\bar{u}} \right) u_0 d \mathbf{r} \\
	\int \left( \ddot{\bar{u}} + \eta \dot{\bar{u}} \right) u_1 d \mathbf{r} \\
	\vdots \\
	\int \left( \ddot{\bar{u}} + \eta \dot{\bar{u}} \right) u_{w-1} d \mathbf{r} \\
	-\int \left\{ \ddot{\bar{u}} + \eta  \dot{\bar{u}} \right\} u  +   \left\{ \ddot{u} + \eta \dot{u} \right\} \bar{u} +   2\dot{\bar{u}} \dot{u}  d \mathbf{r}  \\
\end{bmatrix}
=
\begin{bmatrix}
	- k_0 G_0 \\
  - k_1 G_1 \\
	\vdots \\
	- k_{w-1} G_{w-1} \\
	- k_w G_w \\
\end{bmatrix}
. \label{app_A_G_dyn_eq}
\end{equation} 
From Eq.~(\ref{DFPMVg}) we now have 
\begin{equation}
 \ddot{ \bar{u} } + \eta \dot{ \bar{u }} =  - \hat{H}\bar{u}  - \sum_{j=0}^{w-1} \mu_j \bar{u}_j  +  \mu_w \bar{u}  ,
\end{equation}
with $\hat{H} = - \frac{\hbar^2}{2M} \nabla^2  + V(\mathbf{r}) $ as defined in Eq.~(\ref{eq:TDSE}),
such that the left hand side of Eq.~(\ref{app_A_G_dyn_eq}) is
\begin{equation}
\ddot{\vec{G}} + \eta \dot{\vec{G}}= 
\begin{bmatrix}
	-\int u_0\hat{H}\bar{u} d \mathbf{r} +  \mu_w \int \bar{u}u_0 d \mathbf{r} - \sum_{j=0}^{w-1} \mu_j \int \bar{u}_j u_0   d \mathbf{r} \\
	-\int u_1\hat{H}\bar{u} d \mathbf{r} +  \mu_w \int \bar{u}u_1 d \mathbf{r} - \sum_{j=0}^{w-1} \mu_j \int \bar{u}_j u_1   d \mathbf{r} \\
	\vdots \\
	-\int u_{w-1}\hat{H}\bar{u} d \mathbf{r} +  \mu_w \int \bar{u}u_{w-1} d \mathbf{r} - \sum_{j=0}^{w-1} \mu_j \int \bar{u}_j u_{w-1}   d \mathbf{r} \\
	2E  +  2\mu_w \left(G_w - 1 \right)   +  \sum_{j=0}^{w-1} \mu_j \int \left( \bar{u}_j u + \bar{u} u_j \right) d \mathbf{r}  -  2 \int \dot{\bar{u}} \dot{u}  d \mathbf{r}  \\
\end{bmatrix}
. 
\end{equation} 
Now since $ \int \bar{u}_i u_j d \mathbf{r} = \delta_{ij}$ we can write Eq.~(\ref{app_A_G_dyn_eq}) on matrix form with the Lagrange multipliers as the unknows
\begin{equation}
\begin{bmatrix}
    1 & 	0 &  \hdots & 0 &    -G_0\\
 0 &   1 & 	 \hdots & 0 &    -G_1  \\
	\vdots    &	\vdots  &	& \vdots  &	\vdots \\
0 & 0 &  \hdots &   1   &   -G_{w-1}\\
	     -\operatorname{Re} \left( G_0 \right)  &     -\operatorname{Re} \left( G_1 \right)  & \hdots &  -\operatorname{Re} \left( G_{w-1} \right) &   1-G_{w}    \\
\end{bmatrix}
\begin{bmatrix}
 \mu_0 \\
 \mu_1 \\
	\vdots \\
 \mu_{w-1}  \\
\mu_w   \\
\end{bmatrix}
=
\begin{bmatrix}
	k_0 G_0 - \int u_0\hat{H}\bar{u} d \mathbf{r} \\
	k_1 G_1 - \int u_1\hat{H}\bar{u} d \mathbf{r} \\
	\vdots \\
	k_{w-1} G_{w-1} - \int u_{w-1}\hat{H}\bar{u} d \mathbf{r} \\
	k_{w} G_{w}/2 + E    -   \int | \dot{u} |^2 d \mathbf{r} \\
\end{bmatrix}
, \label{app_A_lin_sys}
\end{equation} 
where $\operatorname{Re} \left( G_j \right) =  \int \left( \bar{u}_j u + \bar{u} u_j \right) d \mathbf{r} /2$ and $E=\int \bar{u} \hat{H} u \, d \mathbf{r} $.

For example with only one ($w=1$) orthogonality constraint, Eq.~(\ref{app_A_lin_sys}) is the system in Eq.~(\ref{app_A_mu_1_and_mu_2}).

We note that the system~(\ref{app_A_lin_sys}) can be solved very efficiently by sparse Gaussian elimination with a computational cost proportional to $w$.

\section{Normalization and angular momentum constraints} \label{Sec_appendix_B}

The two constraints we used for the NLSE are defined in Eq.~(\ref{constraints_for_NLSE}).
Taking the first and second order derivatives of $G_1$ and $G_2$ with respect to $\tau$ gives
\begin{equation}
\dot{G}_1= - \int\limits_{-\pi}^{\pi} \left( \dot{ \bar{u} } u +  \bar{u} \dot{u} \right) dx  , \ \ddot{G}_1=  - \int\limits_{-\pi}^{\pi} \left( \ddot{ \bar{u} } u  +  2 \dot{ \bar{u} } \dot{u}  +   \bar{u} \ddot{u} \right) dx, \label{dot_G_1_NLSE}
\end{equation}
respectively
\begin{equation}
\dot{G}_2=  \mathrm{i} \int\limits_{-\pi}^{\pi} \left( \dot{ \bar{u} } \frac{\partial u}{\partial x} +  \bar{u} \frac{\partial \dot{u}}{\partial x}  \right) dx, 
\ \ddot{G}_2= \mathrm{i} \int\limits_{-\pi}^{\pi} \left( \ddot{ \bar{u} } \frac{\partial u}{\partial x}  + 2 \dot{ \bar{u} } \frac{\partial \dot{u}}{\partial x}    +  \bar{u} \frac{\partial \ddot{u}}{\partial x} \right) dx . \label{dot_G_2_NLSE}
\end{equation}
Inserting the expressions from Eqs.~(\ref{dot_G_1_NLSE}) and~(\ref{dot_G_2_NLSE}) into the left hand side of the general differential equation for constraints~(\ref{gdamp}), then gives after simplifications
\begin{equation}
\ddot{G}_1 + \eta \dot{G}_1 =  - \int\limits_{-\pi}^{\pi} \left( \left\{ \ddot{ \bar{u} } + \eta \dot{ \bar{u} } \right\} u +  \bar{u}  \left\{ \ddot{ u } + \eta \dot{ u } \right\}  + 2 |\dot{u}|^2 \right) dx,  
\end{equation}
and
\begin{equation}
\ddot{G}_2 + \eta \dot{G}_2 =  \mathrm{i} \int\limits_{-\pi}^{\pi} \left( \left\{ \ddot{ \bar{u} } + \eta \dot{ \bar{u} } \right\} \frac{\partial u}{\partial x} +  \bar{u}  \left\{ \frac{\partial \ddot{u}}{\partial x} + \eta \frac{\partial \dot{u}}{\partial x} \right\}  + 2 \dot{ \bar{u} } \frac{\partial \dot{u}}{\partial x} \right) dx.  
\end{equation}
The use of Eq.~(\ref{DE_for_NLSE_with_constraints_for_NLSE}) for the curly brackets above gives (with $\eta, \gamma, \mu$ and $\Omega$ real, and by using integration by parts to some terms)
\begin{equation}
\ddot{G}_1 + \eta \dot{G}_1 = \int\limits_{-\pi}^{\pi} \left( -2 \bar{u} \frac{\partial^2 u}{\partial x^2} + 4\pi \gamma |u|^4  +  2 \mathrm{i} \Omega \bar{u}  \frac{\partial u}{\partial x}  - 2\mu |u|^2 - 2 |\dot{u}|^2 \right) dx, \label{eq_App_B_5}
\end{equation}
and
\begin{equation}
\ddot{G}_2 + \eta \dot{G}_2 =  \int\limits_{-\pi}^{\pi} \left( 2 \mathrm{i}  \frac{\partial u}{\partial x}  \frac{\partial^2 \bar{u}}{\partial x^2} - 2 \mathrm{i} \pi \gamma |u|^2 \bar{u} \frac{\partial u}{\partial x}   - 2 \mathrm{i}    \pi \gamma  \bar{u} \frac{\partial \left( |u|^2 u \right)}{\partial x} + 2 \Omega \bar{u} \frac{\partial^2 u}{\partial x^2}   + 2 \mathrm{i}\mu \bar{u} \frac{\partial u}{\partial x} + 2 \mathrm{i} \dot{ \bar{u} } \frac{\partial \dot{u}}{\partial x} \right) dx. \label{eq_App_B_6}
\end{equation}
Comparing the above equations with Eq.~(\ref{gdamp}) and inserting the constraints~(\ref{constraints_for_NLSE}) 
\begin{equation}
G_{1} = 1 - \int\limits_{-\pi}^{\pi}\left|u\right|^{2}dx \equiv 1 - N   = 0,\ G_{2} = \ell_0 + \mathrm{i}
\int\limits_{-\pi}^{\pi}\bar{{u}} \frac{\partial u}{\partial x} dx  \equiv  \ell_0 - \ell  =0, 
\end{equation}
with the three real functions $N \left( \tau \right), \ \ell \left( \tau \right)$ and $\langle \hat{\ell}^2 \rangle \left( \tau \right) = - \int_{-\pi}^\pi \bar{u} \frac{\partial^2 u}{\partial x^2} dx$, being the norm, the angular momentum, and (here) the kinetic energy, respectively, we have from Eq.~(\ref{eq_App_B_5})  
\begin{equation}
 -2 N \mu  - 2 \ell \Omega    + \int\limits_{-\pi}^{\pi} \left( -2 \bar{u} \frac{\partial^2 u}{\partial x^2} + 4\pi \gamma |u|^4   - 2 |\dot{u}|^2 \right) dx = - k_1 G_1,  \label{CEq1}
\end{equation}
and from Eq.~(\ref{eq_App_B_6}) 
\begin{equation}
  -2 \ell \mu - 2 \langle \hat{\ell}^2 \rangle \Omega  + \int\limits_{-\pi}^{\pi} \left( 2 \mathrm{i} \frac{\partial u}{\partial x} \frac{\partial^2 \bar{u}}{\partial x^2}  - 4 \mathrm{i} \pi \gamma  \frac{\partial u}{\partial x}  |u|^2 \bar{u}   - 2 \mathrm{i} \pi \gamma   \frac{\partial \left( \bar{u} u \right)}{\partial x} \bar{u} u   + 2 \mathrm{i} \dot{ \bar{u} } \frac{\partial \dot{u}}{\partial x}  \right) dx = - k_2 G_2.  \label{CEq2}
\end{equation}
The second to last term in the left hand side above disappears, since $\int \frac{\partial \left( \bar{u} u \right)}{\partial x} \bar{u} u dx =  \int \frac{\partial}{\partial x} \left( \bar{u} u \right)^2  dx /2=  \left(  |u \left( \pi \right) |^4 - |u \left( -\pi \right) |^4 \right) /2 = 0$ due to the periodic boundary conditions.
Hence, Eqs.~(\ref{CEq1}) and~(\ref{CEq2}) leads us to the following linear system for the Lagrange multipliers
\begin{equation}
\begin{bmatrix}
	N & \ell \\
	\ell & \langle \hat{\ell}^2 \rangle \\
\end{bmatrix}
\begin{bmatrix}
	\mu \\
	\Omega \\
\end{bmatrix}
=
\begin{bmatrix}
	 k_1 G_1/2 + \int_{-\pi}^\pi \bar{u} \hat{H} u \, dx  - \int_{-\pi}^\pi | \dot{u} |^2 dx \\ 
	 k_2 G_2/2 - \mathrm{i} \int_{-\pi}^\pi \frac{\partial u}{\partial x} \hat{H} \bar{u} \, dx + \mathrm{i} \int_{-\pi}^\pi  \bar{ \dot{u} }   \frac{\partial \dot{u}}{\partial x} dx \\
\end{bmatrix}
\equiv
\begin{bmatrix}
	b_1 \\
	b_2 \\
\end{bmatrix}
, \label{eq_app_B_linear_system_NLSE}
\end{equation}
where $\hat{H} u  = \delta E / \delta \bar{u}$ with $E$ from Eq.~(\ref{eq:NLSE_total_energy}).
Using Cramer's rule on the above linear system gives the explicit expressions used in Eq.~(\ref{LM_for_NLSE_with_constraints_for_NLSE}).

\vspace{10mm} 

\end{document}